\documentclass[]{pasj01}

\Received{}
\Accepted{}

\newcommand{\vmax}{V_\mathrm{max}}
\newcommand{\rmax}{r_\mathrm{max}}

\newcommand{\rN}{r_\mathrm{N}}
\newcommand{\rhoN}{\rho_\mathrm{N}}
\newcommand{\fN}{f_\mathrm{N}}
\newcommand{\xN}{x_\mathrm{N}}
\newcommand{\rB}{r_\mathrm{B}}
\newcommand{\rhoB}{\rho_\mathrm{B}}
\newcommand{\fB}{f_\mathrm{B}}
\newcommand{\xB}{x_\mathrm{B}}
\newcommand{\rhocrit}{\rho_\mathrm{crit, 0}}

\newcommand{\ce}[1]{$\mathrm{#1}$}

\begin{document} 

\title{
A universal scaling relation incorporating the cusp-to-core transition of dark matter haloes
}

\author{Yuka \textsc{Kaneda}\altaffilmark{1}%
\thanks{Example: Present Address is xxxxxxxxxx}}
\altaffiltext{1}{Graduate School of Science and Technology, University of Tsukuba, 1-1-1 Tennodai, Tsukuba, Ibaraki, 305-8577, Japan}
\email{kaneda@ccs.tsukuba.ac.jp}

\author{Masao \textsc{Mori}\altaffilmark{2}}
\altaffiltext{2}{Center for Computational Sciences
, University of Tsukuba, 1-1-1 Tennodai, Tsukuba, Ibaraki, 305-8577, Japan}

\author{Koki \textsc{Otaki},\altaffilmark{3}}
\altaffiltext{3}{Amanogawa Galaxy Astronomy Research Center, Graduate School of Science and Engineering, Kagoshima University, 1-21-35 Korimoto, Kagoshima, Kagoshima 890-0065, Japan}



\maketitle

\begin{abstract}
The dark matter haloes associated with galaxies have hitherto established strong correlations within a range of observed parameters, known as scaling relations of dark matter haloes. 
The origin of these scaling relations still contains significant ambiguities and requires comprehensive exploration for complete understanding.
Utilising the correlation between the concentration and mass of dark matter haloes inferred from cosmological $N$--body simulations based on the cold dark matter paradigm ($c$--$M$ relation), we derive theoretical scaling relations among other physical quantities such as the surface mass density, the maximum circular velocity, and the scale radius of the dark matter halo. 
By comparing theoretical and observed scaling relations at various mass scales, it is found that the scaling relations observed in dwarf galaxies and galaxies originate in the $c$--$M$ relation of the dark matter halo.
We predict that this theoretical scaling relation is also established in galaxy clusters. 
Moreover, we propose a novel theoretical scaling relation that incorporates the effects of the cusp--to--core transition, which is supposed to occur in cold dark matter haloes. 
Our discussion concludes with the exploration of potential observational verification of the cusp--to--core transition process in dark matter haloes.
\end{abstract}


\section{Introduction} \label{sec:intro}

Ninety years have passed since \citet{zwicky_rotverschiebung_1933} postulated the existence of a mysterious 'missing mass' in galaxy clusters, later referred to as dark matter.
Since then, the evidence for dark matter has emerged from a variety of astrophysical observations, including the rotation curves of galaxies and the large-scale structure of the universe. These observations suggest that there is more mass in the universe than can be accounted for by visible matter.
The nature of dark matter remains one of the most profound mysteries in astrophysics and particle physics.

From an observational point of view, the scaling relations among dark matter haloes play a vital role in exploring the nature of dark matter.
The scaling relations of dark matter haloes refer to the empirical relationships or correlations between various properties of the cosmic structures.
These properties often include mass, size, concentration, and other characteristics of the dark matter haloes.
So far, they have been intensively explored, coupled with the fact that they provide strong constraints on galaxy formation.
In particular, \citet{burkert_structure_1995} found a correlation between characteristic radii and circular velocities at the characteristic radii of dark matter haloes associated with galaxies based on the observed data of seven dwarf galaxies.
Following this, some studies have pointed out that the central surface density of a dark matter halo, represented by the production of the central density and the core radius, is constant over a wide range with respect to the absolute $B$-band magnitude of nearby galaxies \citep{kormendy_scaling_2004, spano_ghasp_2008, donato_constant_2009, salucci_dwarf_2012, kormendy_scaling_2016, salucci_distribution_2019}.
On the other hand, recent studies \citep{hayashi_universal_2017, li_constant_2019, del_popolo_sparc_2023, gopika_test_2023} have claimed that the central surface density weakly depends on the scale of galaxies.
In any case, these correlations should reflect the dynamical processes which drive the formation and evolution of galaxies. 
However, the origin of these empirical relations remains an open question.

The $\Lambda$ cold dark matter (CDM) theory has demonstrated exceptional efficacy in elucidating and accounting for many of observational phenomena on scales larger than galaxies.
However, this has been considered to encounter significant challenges when studying the galactic and sub-galactic scales.
The cusp--core problem \citep{moore_nature_1994} is one of the notable issues.
Cosmological $N$--body simulations based on the CDM theory yield a universal mass density distribution that diverges at the centre of dark matter haloes
(e.g. \cite{navarro_simulations_1995}; \cite{navarro_structure_1996}; \cite{fukushige_origin_1997}; \cite{navarro_universal_1997}; \citet{moore_cold_1999}; \cite{navarro_diversity_2010}; \cite{ishiyama_cosmogrid_2013}).
The Navarro--Frenk--White (NFW) profile \citep{navarro_universal_1997} fits well the density profiles of dark matter haloes formed in CDM simulations.
In addition, many past studies claim that the density distributions of haloes formed in cosmological CDM simulations have a tight correlation between their concentrations, their virial masses, and the redshift $z$, called concentration--mass relation ($c$--$M$ relation, \cite{prada_halo_2012}; \cite{klypin_multidark_2016}).
This relation is an essential tool for characterising the mass density distribution of dark matter haloes formed in CDM simulations.

On the other hand, observations of low--mass galaxies indicate that a substantial fraction of their dark matter haloes have constant density at small radii, a distribution called core profile (e.g. \cite{moore_nature_1994}; \cite{burkert_structure_1995}; \cite{de_blok_high-resolution_2001}; \cite{de_blok_mass_2001}; \cite{oh_high-resolution_2008}; \cite{oh_central_2011}; \cite{oh_high-resolution_2015}).
The Burkert profile \citep{burkert_structure_1995} fits well the density structure of the observed dark matter haloes with cores.
This long--standing discrepancy between simulated and observed central density distributions of the dark matter haloes is commonly referred to as the 'cusp--core problem', and has been the subject of intensive analysis.

Several explanations of this discrepancy have been proposed.
One of them is to consider dark matter particles which intrinsically produce a core, such as self--interactive dark matter \citep{spergel_observational_2000} and fuzzy dark matter \citep{weinberg_new_1978, hui_ultralight_2017, burkert_fuzzy_2020}.
The other is a baryonic solution within the CDM framework.
The mechanisms that transform a central cusp into a core have been actively investigated, taking into account the effect of baryon feedback, which is not included in the cosmological $N$--body dark matter only simulations. 
One mechanism is the dynamical friction, caused by infalling massive gas clumps, heats up dark matter, and the distribution of dark matter expands, leading to diffuse distribution at the centre (e.g. \cite{el-zant_dark_2001}; \cite{goerdt_core_2010}; \cite{inoue_cores_2011}).
Another and most promising one is the stellar feedback, such as supernova--driven outflow that alters the gravitational potential of a system (\cite{navarro_cores_1996}; \cite{gnedin_maximum_2002}; \cite{read_mass_2005}; \cite{ogiya_core-cusp_2011}; \cite{ragone-figueroa_effects_2012}; \cite{ogiya_core-cusp_2014}; for review see, e.g. \cite{bullock_small-scale_2017}; \cite{salucci_distribution_2019}; \cite{del_popolo_review_2022}).
In the following, we refer to these baryonic effects that alter a cusp into a core as the "cusp--to--core transition" mechanism.

In the stellar feedback model, \citet{ogiya_core-cusp_2011} suggested that only one cycle of stellar feedback is not enough to change the density distribution of a whole system irreversibly.
\citet{ogiya_core-cusp_2014} showed that recurrent change of potential arising from the oscillatory feedback process is the primary driving mechanism of the cusp--to--core transition which generates a dynamically stable core.
Furthermore, they found a simple relationship between the oscillation period of the stellar feedback and the core radius of the dark matter haloes assuming that the energy transport process is supported by Landau resonance.

While \citet{ogiya_core-cusp_2014} employed an idealised set-up to investigate the effect of potential oscillation, the more realistic effects of stellar feedback are investigated utilising zoom--in cosmological hydrodynamic simulations (e.g. \cite{madau_dark_2014}; \cite{fitts_fire_2017}; \cite{benitez-llambay_baryon-induced_2019}; \cite{lazar_dark_2020}).
One of the important findings is that the core formation depends on the amount of formed stars in one system. 
When a galaxy forms a sufficient number of stars, the energy produced by supernovae feedback is enough to redistribute dark matter and create a large core. 
On the other hand, in the system with the ratio of stellar mass to virial mass less than $10^{-4}$, the effect of baryon feedback is negligible and a cusp remains (\cite{di_cintio_dependence_2014}; \cite{tollet_nihao_2016}; \cite{lazar_dark_2020}). 
However, the approximations used to implement the star--formation and feedback in simulations have a significant impact on core formation.
For example, individual supernova remnants remain unresolved because their size ($\sim 10 \, \mathrm{pc}$) is comparable to the resolution of simulations. 
The densities of molecular clouds ($\sim 10^{3-4} \, \mathrm{cm^{-3}}$) are unresolved and the simulations adopt artificial density thresholds (e.g. $\sim 100 \, \mathrm{cm^{-3}}$) for star formation. 
Therefore, more sophisticated methods for including star formation and stellar feedback in cosmological simulations as sub-grid models are expected.

This study aims to investigate the origins of scaling relations in dark matter haloes and assess the impact of the cusp--to--core transition on these relations.
We seamlessly integrate the scaling relation from dwarf galaxies to clusters of galaxies introducing a simple and reasonable cusp--to--core transition model that remains independent of the type of baryonic feedback. 
Subsequently, the range of system mass in which the cusp--to--core transition occurs is discussed. 
Furthermore, we evaluate how this process and the critical mass for the transition can be observed using the latest and upcoming facilities and instruments.

This paper is organised as follows: 
In section \ref{sec:parameters}, the profiles and the parameters of dark matter haloes used in this paper are summarised.
In section \ref{sec:scaling}, we relate the origin of the scaling relation for galaxy-scale dark matter haloes to the $c$--$M$ relation.  
Then, we further investigate the scalings for MW satellites in detail in section \ref{sec:MWdwarf}.  
In section \ref{sec:wide}, targets are extended to groups and clusters of galaxies. 
We investigate whether the scaling relations originating from the $c$--$M$ relation are reproduced in a wide mass range from dwarf galaxies to clusters of galaxies.
Until section \ref{sec:wide}, we use a cuspy profile, which is the density distribution of haloes typically arising in cosmological $N$--body simulations. 
However, cored profiles more accurately replicate the observed density distributions.
Therefore, we model a process that transforms a cusp profile into a core profile in order to derive the parameters of a cored profile, such as core radius and central density, after the cusp--to-core transition, assuming the parameters of a cusp profile according to the $c$--$M$ relation. In section \ref{sec:CCtrans}, we predict a scaling relation for cored haloes using this model.
In section \ref{sec:discussion}, we discuss the capability of the new scaling relation and application of the cusp--to--core transition model.
To conclude, our work is summarised in section \ref{sec:conclusion}.

\section{Definition of the dark matter halo properties} \label{sec:parameters}

The density distribution of a CDM halo is well reproduced by the NFW profile: 
\begin{equation}
    \rho(r)=\frac{\rhoN}{\frac{r}{\rN}\left(\frac{r}{\rN}+1\right)^2}, \label{eq:rho_nfw}
\end{equation}
where $r$ is the distance from the centre and $\rhoN$ and $\rN$ represent the scale density and scale radius of a halo, respectively.
The mass enclosed within the radius $r$ can be derived by integrating equation (\ref{eq:rho_nfw}) as
\begin{equation}
    M(<r) = 4\pi\rho_\mathrm{N} r_\mathrm{N}^3 f_\mathrm{N}\left(\frac{r}{r_\mathrm{N}}\right) \label{eq:m_nfw},
\end{equation}
where
\begin{equation}
    f_\mathrm{N}(x)=\ln \left(1+x\right)-\frac{x}{1+x}. \label{eq:f_nfw}
\end{equation}
Many previous studies (see section \ref{sec:intro}) have claimed that these haloes with central cusps formed in cosmological CDM simulations present a tight correlation between their concentrations of mass $c_{200}$, their masses $M_{200}$, and the redshift $z$, called the $c$--$M$ relation.
$c_{200}$ is defined by
\begin{equation}
    c_{200}\equiv\frac{r_{200}}{r_\mathrm{N}}.  \label{eq:c200}
\end{equation}
The relational expression between the virial mass $M_{200}$ and the virial radius $r_{200}$ for $z = 0$ can be written by
\begin{equation}
     M_{200}\equiv\frac{4}{3}\pi200\rho_{\mathrm{crit,0}}r_{200}^3. \label{eq:m200}
\end{equation}
$\rho_{\mathrm{crit},0} = 3H_0^2/(8\pi G)$ is the critical density of the universe at $z = 0$.
$H_0$ is the Hubble constant and can be expressed as $H_0 = 100 \, h \, \rm{km \, s^{-1} \, Mpc}$ using the dimensionless Hubble constant $h = 0.674$ \citep{planck_collaboration_planck_2020}.

The circular velocity is defined as
\begin{equation}
   V(r) = \left( \frac{GM(<r)}{r} \right)^{1/2}
\end{equation}
Then the maximum circular velocity $\vmax$ is given by $\vmax = V(\rmax)$ with the radius at which the circular velocity becomes maximum, $r_\mathrm{max}$.
$\rmax$ can be derived by solving $dV/dr=0$.
Using the NFW profile as a mass distribution, we obtain
\begin{equation}
   \rmax=\xN \, \rN \label{eq:rmaxN}
\end{equation}
where $\xN = 2.16$. 

There are two methods to estimate halo concentration from a resulting distribution of particles in $N$--body simulations. 
Firstly, one can calculate the spherically averaged density distribution and then fit it with an appropriate function. 
In this case, the results may depend upon the selected binning algorithm of particle position and the fit quality. 
Secondary, instead of making fits to individual halo profiles, one can find the concentration by solving the following equations numerically, having the $\vmax/V_{200}$ ratio for each halo \citep{klypin_dark_2011, prada_halo_2012}:
\begin{equation}
   \frac{V_\mathrm{max}}{V_{200}}=\left(\frac{c_{200}\fN(\xN)}{\xN \fN(c_{200})}\right)^{1/2}. \label{eq:v200_vmax}
\end{equation}
Here, we regard the virial velocity $V_{200}$ to be $V_{200} = (GM_{200}/r_{200})^{1/2}$.
In this paper, we use the concentrations found by the later method.

The Burkert profile \citep{burkert_structure_1995} fits well the density structures of dark matter haloes with cores,
\begin{equation}
    \rho(r) = \frac{\rho_\mathrm{B}}{\left(\frac{r}{r_\mathrm{B}}+1\right)\left(\frac{r^2}{{r_\mathrm{B}}^2}+1\right)}, \label{eq:rho_burkert}
\end{equation}
where $\rhoB$ and $\rB$ are a halo's central density and core radius, respectively.
The mass enclosed within the radius $r$ can be derived by integrating equation (\ref{eq:rho_burkert}) as
\begin{equation}
    M(<r) = 4 \pi \rho_\mathrm{B} r_\mathrm{B}^{3} f_\mathrm{B}\left(\frac{r}{r_\mathrm{B}}\right) \label{eq:m_burkert},
\end{equation}
where
\begin{equation}
    f_\mathrm{B}(x) =\frac{1}{4} \ln \left(x^{2}+1\right)+\frac{1}{2} \ln (x+1)-\frac{1}{2} \arctan x. \label{eq:f_burkert}
\end{equation}
$\rmax$ for the Burkert profile can also be derived by solving $dV/dr=0$ which gives
\begin{equation}
   \rmax=\xB \, \rB,
\end{equation}
where $\xB = 3.24$.

The pseudo--isothermal sphere (p-ISO) can also represent a cored profile, 
\begin{equation}
    \rho(r) = \frac{\rho_\mathrm{p-ISO}}{1+\frac{r^2}{r_\mathrm{p-ISO}^2}}.
\end{equation}
where $r_\mathrm{p-ISO}$ is the core radius of p-ISO and $\rho_\mathrm{p-IS0}$ is the central density.
The mass enclosed within the radius $r$ can be written as
\begin{equation}
    M(<r) = 4 \pi \rho_\mathrm{p-ISO} r_\mathrm{p-ISO}^{3} f_\mathrm{p-ISO}\left(\frac{r}{r_\mathrm{p-ISO}}\right) \label{eq:m_p-ISO},
\end{equation}
where
\begin{equation}
    f_\mathrm{p-ISO}(x) = x - \arctan x. \label{eq:f_p-ISO}
\end{equation}

\section{Origin of the observed scaling relations} \label{sec:scaling}

\begin{table*}
 \tbl{Original scaling relations and the relations converted to $\bar{\Sigma}(<\rmax)$ as a function of $\vmax$.}{%
\begin{tabular}{lccccc}
\hline
Name & \citet{burkert_structure_1995} & \citet{spano_ghasp_2008} \\
\hline
original & $\rho_\mathrm{B} = 4.5\times10^{-2} (r_\mathrm{B}/\mathrm{kpc})^{-2/3} \, M_\odot\,\mathrm{pc}^{-3}$ & 
$\rho_\mathrm{p-ISO}r_\mathrm{p-ISO} \sim 150 \, M_\odot \mathrm{pc}^{-2}$  \\
converted to $\bar{\Sigma}(<\rmax, \vmax)$ [$*$] & $\bar{\Sigma}(<\rmax) = 2.5 \{\vmax/(\mathrm{km \, s}^{-1})\}^{1/2} M_\odot \, \mathrm{pc}^{-2}$ & $\bar{\Sigma}(<\rmax) \sim 80 \, M_\odot \, \mathrm{pc}^{-2}$ \footnotemark[$*$]\\    
\hline \\
\hline
\citet{donato_constant_2009} & \citet{donato_constant_2009} dwarf & \citet{kormendy_scaling_2016}\\
\hline
$\rhoB \rB  = 140^{+80}_{-30} \, M_\odot \mathrm{pc}^{-2}$ & 
$\rhoB \rB \sim 100 \, M_\odot \mathrm{pc}^{-2}$ & 
$\rho_\mathrm{NI}r_\mathrm{NI} = 70\pm4 \, M_\odot \, \mathrm{pc}^{-2}$ \\
$\bar{\Sigma}(<\rmax) = 37^{+21}_{-8} \, M_\odot \, \mathrm{pc}^{-2}$ & 
$\bar{\Sigma}(<\rmax) \sim 27 \, M_\odot \, \mathrm{pc}^{-2}$ & 
$\bar{\Sigma}(<\rmax) = 38\pm2 \, M_\odot \, \mathrm{pc}^{-2}$ \footnotemark[$\dag$]\\
\hline
\end{tabular}} \label{tab:scaling}
\begin{tabnote}
\footnotemark[$*$] The parameters of p-ISO are converted to the ones of the Burkert profile utilising the method described in appendix \ref{app:p-ISO_to_Burkert} since p-ISO profile do not have maximum circular velocity. \\
\footnotemark[$\dag$] The parameters of the non-singular isothermal sphere are considered to be the same as the parameters of p-ISO.
\end{tabnote}
\end{table*}

\begin{figure*}
 \begin{center}
  \includegraphics[width=\hsize]{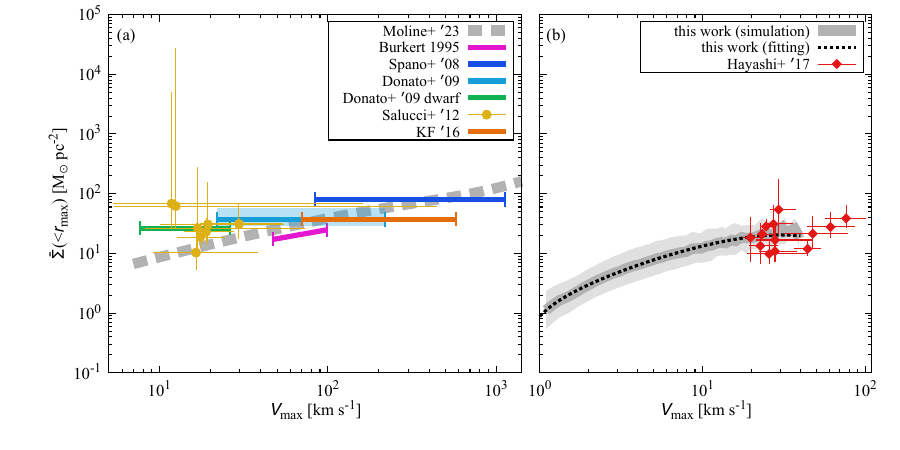}
 \end{center}
\caption{Left panel: a comparison between the scaling relations proposed by \citet{burkert_structure_1995} (magenta line),  \citet{spano_ghasp_2008} (blue line), \citet{donato_constant_2009} (sky--blue line for galaxies, green line for dwarf galaxies), \citet{salucci_dwarf_2012} (yellow symbol), \citet{kormendy_scaling_2016} (orange line) and the theoretical $\bar{\Sigma}(<\rmax)$--$\vmax$ relation by \citet{moline_cdm_2022} (grey line).
The coloured line segments start and end with the domains of definitions of each scaling relation.
The propagated errors are shown in the shaded region in sky-blue \citep{donato_constant_2009} and 
in orange, which is narrower than the thickness of the solid line \citep{kormendy_scaling_2016}.
As a theoretical $\bar{\Sigma}(<\rmax)$--$\vmax$ relation, we adopt the fitting curve of the $c_V$--$\vmax$ relation of subhaloes from the Uchuu suite \citep{ishiyama_Uchuu_2021} for $z=0$ proposed by \citet{moline_cdm_2022} (equation \ref{eq:cv-Vmaxrel} with $C=c_{\mathrm{V}}(V_{\max })$ and $S=V_{\max } \, \mathrm{km}\, \mathrm{s}^{-1}$, and the coefficients shown in equation \ref{eq:cv-Vmaxrel_Moline2023}) converted to the $\bar{\Sigma}(<\rmax)$--$\vmax$ plane using equation (\ref{eq:Sigmavmax-Vmax_relation}).
Right panel: the median of $\bar{\Sigma}(<\rmax)$ against $\vmax$ derived for subhaloes of MW--like hosts in the high-resolution cosmological $N$--body simulation, Phi-4096 (\citet{ishiyama_Uchuu_2021}, the thick solid line).
The fitting curve of the $c_V$--$\vmax$ relation of MW subhaloes from Phi-4096 for $z=0$ (equation \ref{eq:cv-Vmaxrel} with $C=c_{\mathrm{V}}(V_{\max })$ and $S=V_{\max } \, \mathrm{km}\, \mathrm{s}^{-1}$, and the coefficients shown in equation \ref{eq:cv-Vmaxrel_MWsat}) converted to the $\bar{\Sigma}(<\rmax)$--$\vmax$ plane using equation (\ref{eq:Sigmavmax-Vmax_relation}) is shown in black dotted line.
The shaded region shows a 25-75th percentile around it. 
The red symbols show observed MW and M31 satellites. 
}
\label{fig:scalings}
\end{figure*}

In this section, we attribute the origin of the empirical scaling relation for dark matter haloes introduced in section \ref{sec:intro} to the theoretical universality, the $c$--$M$ relation, derived from cosmological dark matter only simulations.
We transform the empirical scaling relation into equations on the surface density and $V_{\rm max}$ plane. 
Similarly, the $c$--$M$ relations by \citet{moline_cdm_2022}, derived from the recent cosmological $N$--body simulation \citep{ishiyama_Uchuu_2021}, is projected into relationships between surface density and $V_{\rm max}$. 
The mutual connection between the empirical and theoretical scaling relation, which govern the parameters of dark matter haloes, is subsequently examined.

\citet{burkert_structure_1995} suggested the existence of a simple correlation between the core radius of the Burkert profile $r_\mathrm{B}$ and the central density $\rho_\mathrm{B}$.
This relation is based on seven nearby dwarf galaxies which are clearly dark matter dominated at their core radius.
The parameters of the dark matter haloes are determined by fitting the theoretically predicted rotation curve to the observational \ce{H_I} rotation curves.
They derived the following correlation:
$\rho_\mathrm{B} = 4.5\times10^{-2} (r_\mathrm{B}\;\mathrm{kpc}^{-1})^{-2/3}M_\odot\,\mathrm{pc}^{-3}$.

\citet{kormendy_scaling_2004} firstly pointed out that the central surface density, the product of a central density and a core radius is constant and independent of galaxy luminosity.
This relation was updated in \citet{kormendy_scaling_2016}.
\citet{kormendy_scaling_2016} claimed that central surface densities of dark matter haloes of Sc-Im galaxies and dwarf galaxies are constant independent of their absolute $B$-band magnitude, 
$\rho_\mathrm{NI}r_\mathrm{NI}=70\pm4\,M_\odot\,\mathrm{pc}^{-2}$,
where $\rho_\mathrm{NI}$ is the central density and $r_\mathrm{NI}$ is the scale radius of the nonsingular isothermal sphere.
However, the central surface densities of dwarf galaxies were artificially increased to match their scaling relation with that of disk galaxies.
Therefore we exclude the mass range of dwarf galaxies from the domain of the Kormendy-Freeman relation shown in the orange line in figure \ref{fig:scalings}a.

\citet{spano_ghasp_2008} studied the mass distribution of 36 spiral galaxies based on rotation curve decomposition. 
p-ISO is adopted as a dark matter density distribution.
They suggested nearly constant central surface density around $\sim 150 \, M_\odot \mathrm{pc}^{-2}$ independent of galaxy luminosity.

\citet{donato_constant_2009} extended this finding.
Their sample comprised the rotation curves of $\sim 1000$ spiral galaxies, the mass models of individual dwarf irregular and spiral galaxies of late and early types with high-quality rotation curves, and the galaxy--galaxy weak gravitational lensing signals from a sample of spiral and elliptical galaxies.
They determined that $\rhoB \rB = 140^{+80}_{-30} \, M_\odot \mathrm{pc}^{-2}$ for Burkert profile.
They also showed that the observed kinematics of 6 Local Group dwarf spheroidal galaxies indicate $\rhoB \rB \sim 100 \, M_\odot \mathrm{pc}^{-2}$.

Furthermore, \citet{salucci_dwarf_2012} extended this relation to the MW dSphs. 
They performed Jeans analysis for eight MW dSphs and derived the parameters of the Burkert profiles.
They showed that the MW dSphs lie on the extrapolation of the scaling law seen in spiral galaxies.
Note that they did not fit the eight dwarf galaxies with a function of constant central surface density; just confirmed that the dSphs were consistent with the extrapolation of the spiral galaxy scaling proposed by \citet{donato_constant_2009}.

To compare these observational scaling relations, assuming a core profile, with halo properties from CDM simulations, we analyse the scaling relations through a parameter that can be defined for both core and cusp profiles.
Here, we define the characteristic mean surface density of the dark matter halo within the radius of the maximum circular velocity as follows:
\begin{equation}
      \bar{\Sigma}(<\rmax)=\frac{M(<\rmax)}{\pi\rmax^2}, \label{eq:Sigma_vmax_definition}
\end{equation}
where
\begin{equation}
M(<\rmax)=\int_0^{\rmax}4\pi\rho_\mathrm{dm}\left(r'\right)r'^2dr',
\end{equation}
$\rho_\mathrm{dm}$ 
corresponds to the density profile of a dark matter halo
(see \cite{hayashi_common_2015, hayashi_universal_2017}).
Note that this quantity differs from the usual surface density, i.e., the mass contained in the sphere divided by the area, not the mass contained in the cylinder.
If we assume the NFW profile, the Burkert profile or p-ISO as $\rho_\mathrm{dm}$, $\bar{\Sigma}(<\rmax)$ is proportional to the product of scale density and the scale radius or the central density and the core radius.
This means that $\bar{\Sigma}(<\rmax)$ is several times the central surface density that is reported to be constant for galaxies and dwarf galaxies in the literature.
The original observational scaling relations and the ones converted to the $\vmax-\bar{\Sigma}(< \rmax)$ plane are summarised in table \ref{tab:scaling}.
To derive $\bar{\Sigma}(< \rmax)-\vmax$ relation, the parameters of p-ISO are converted to the ones of the Burkert profile using the method described in appendix \ref{app:p-ISO_to_Burkert} since p-ISO do not have maximum circular velocity. 
The parameters of the non-singular isothermal sphere are considered to be the same as the parameters of p-ISO.
Figure \ref{fig:scalings}a shows the scaling relations given in table \ref{tab:scaling} on the $\vmax-\bar{\Sigma}(< \rmax)$ plane. 
Solid lines with different colours represent each scaling relation. 
The propagated errors are shown in the shaded region in sky-blue \citep{donato_constant_2009} and 
in orange, which is narrower than the thickness of the solid line \citep{kormendy_scaling_2016}.
The properties of dwarf galaxies estimated by \citet{salucci_dwarf_2012} are shown in the yellow symbols.

We explore the connection between these scaling relations derived from observations and the theoretical properties of dark matter haloes. 
Theoretical analyses based on cosmological simulations have widely recognized the existence of universal properties of the cold dark matter haloes. 
They manifest as a strong correlation between the total mass and central concentration of the dark matter halo, often referred to as the $c$--$M$ relation. 
In the following, we will investigate the trajectory of this $c$--$M$ relation on the  $\vmax-\bar{\Sigma}(<\rmax)$ plane.

We adopt the fitting curve of the $c$--$M$ relation of subhaloes from the Uchuu suite \citep{ishiyama_Uchuu_2021} for $z=0$ proposed by \citet{moline_cdm_2022}:
\begin{equation}
    C(S) = c_0\left[1+\sum_{i=1}^3 a_i\left[\log _{10}\left(S\right)\right]^i\right], \label{eq:cv-Vmaxrel}
\end{equation}
where $C=c_{\mathrm{V}}(V_{\max })$, $S=V_{\max } / \mathrm{km}\, \mathrm{s}^{-1}$ and 
\begin{equation}
    c_0=1.75 \times 10^5 \quad \mathrm{and} \quad a_i=\{-0.9,0.27,-0.02\}. \label{eq:cv-Vmaxrel_Moline2023}
\end{equation}
These coefficients are also listed in table \ref{tab:fitting_coefficient_c-M}.
Here, the definition of $c_{\mathrm{V}}$ is
\begin{equation}
    c_{\mathrm{V}}=\frac{\bar{\rho}(<\rmax)}{\rhocrit}=2\left(\frac{\vmax}{H_0 \rmax}\right)^2, \label{eq:cv_definition}
\end{equation}
where $\bar{\rho}(<\rmax)$ is the mean physical density attained within the radius corresponding to the maximum circular velocity, $\rmax$, expressed in units of the critical density of the Universe \citep{diemand_formation_2007, moline_characterization_2017}.
The relation between $c_{\mathrm{V}}$ and $c_{200}$ is given by \citet{diemand_formation_2007},
\begin{equation}
    c_{\mathrm{V}} = 200 \left(\frac{c_{200}}{\xN}\right)^3 \frac{\fN(\xN)}{\fN(c_{200})}.
\end{equation}
The maximum circular velocity is connected with the virial mass by equation (\ref{eq:v200_vmax}).
In this way, a $c$--$M$ relation can be written as a $c_{\rm{V}}$--$\vmax$ relation.
Equation (\ref{eq:cv-Vmaxrel}) with (\ref{eq:cv-Vmaxrel_Moline2023}) is valid for $7 \mathrm{~km} \mathrm{~s}^{-1} \lesssim V_{\max} \lesssim 1500 \mathrm{~km} \mathrm{~s}^{-1}$ due to the limitation of halo masses in the simulation.
The $c_{\rm{V}}$--$\vmax$ relation (equation \ref{eq:cv-Vmaxrel} and \ref{eq:cv-Vmaxrel_Moline2023}) expressed on the $c_{200}-M_{200}$ plane is also described by equation (\ref{eq:cv-Vmaxrel}) with $C=c_{200}(M_{200})$, $S=M_{200} / M_\odot$, and
\begin{equation}
     c_0=48.29 \quad \mathrm{and} \quad a_i=\{-0.01863,-0.008262,0.0003651\} \label{eq:c200-M200_rel_wide}
\end{equation}
The error of this parameterisation is less than 3 per cent. 
The $c_{\rm{V}}$--$\vmax$ relation relation can also be expressed in terms of $\bar\Sigma(<\rmax)$ written as a function of $\vmax$. 
Solving equation (\ref{eq:cv_definition}) for $\rmax$ and substitute it into equation (\ref{eq:Sigma_vmax_definition}) yields
\begin{eqnarray}
    \frac{\bar\Sigma(<\rmax)}{M_\odot \,\mathrm{pc}^{-2}} 
    &= 1.10 \times 10^{-12} \; \frac{h}{\pi} \left(\frac{G}{\mathrm{m^3\, kg^{-1} \, s^{-2}}}\right)^{-1} \nonumber\\
    &\times \frac{\vmax}{\mathrm{km\,s^{-1}}} \left\{c_\mathrm{V}\left(\frac{\vmax}{\mathrm{km\,s^{-1}}}\right)\right\}^{1/2}. \label{eq:Sigmavmax-Vmax_relation}
\end{eqnarray}

The grey dashed curve in figure \ref{fig:scalings}a is the theoretical $\bar{\Sigma}(<\rmax)$ curve drawn on the basis of the $c_{\rm{V}}$--$\vmax$ relation of the CDM haloes in equation (\ref{eq:Sigmavmax-Vmax_relation}).
This function is slowly monotonically increasing on this plane.
The scaling relations and the theoretical curve are consistent for the domain of the definition of each scaling relation.
The relations proposed by \authorcite{donato_constant_2009} (\yearcite{donato_constant_2009}, sky--blue line) and \authorcite{kormendy_scaling_2016} (\yearcite{kormendy_scaling_2016}, orange line) correspond to the intermediate mass range around $\vmax\sim100\,\mathrm{km\,s^{-1}}$ and they coincide with the theoretical $\bar{\Sigma}(<\rmax)$--$\vmax$ relation.
The relation proposed by \authorcite{spano_ghasp_2008} (\yearcite{spano_ghasp_2008}, blue line) shows the largest velocities, from $100$ to $1000 \, \mathrm{km\, s^{-1}}$, and the highest characteristic mean surface density, around $100 \, M_\odot\, \mathrm{pc^{-2}}$.
The relation proposed by \citet{spano_ghasp_2008} is also consistent with the theoretical $\bar{\Sigma}(<\rmax)$--$\vmax$ relation.
The empirical scaling relations seem to be broken fragments of the theoretical $\bar{\Sigma}(<\rmax)$--$\vmax$ relation which is derived from $c_{\rm{V}}$--$\vmax$ relation.
In other words, what we have thought of as empirical scaling relations between parameters of the dark matter haloes is, actually, the $c$--$M$ relation, which roots in the nature of the CDM haloes,  seen from a different perspective.
Observations over a limited and narrow mass range make the observed scaling relations appear constant, just as the theoretical $\bar{\Sigma}(<\rmax)$--$\vmax$ relation predicts.
However, the theoretical $\bar{\Sigma}(<\rmax)$--$\vmax$ relation decreases as the mass of the dark matter halo decreases.
Therefore, the arguments that the surface density depends on the scale of a galaxy \citep{hayashi_universal_2017, li_constant_2019, del_popolo_sparc_2023, gopika_test_2023} can be understood as a natural consequence of the $c$--$M$ relation.

The scaling relation of dwarf galaxies has been poorly discussed in previous studies.
\authorcite{donato_constant_2009} (\yearcite{donato_constant_2009}, green line) fitted only six dwarf galaxies with a constant.
\authorcite{salucci_dwarf_2012} (\yearcite{salucci_dwarf_2012}, yellow symbol) increases the number of dwarf galaxies to eight, however, they have too large uncertainties to limit the scaling relation of dwarf galaxies. 
In the next section, we discuss the scaling relation of dwarf galaxies ($V_\mathrm{max} \lesssim 100\,\mathrm{km\,s^{-1}}$) based on recent estimations with improved observation.

\begin{table*}
  \tbl{Coefficients of equation (\ref{eq:cv-Vmaxrel}).\footnotemark[$*$] }{%
  \begin{tabular}{ccccccccc}
\hline 
$C$ & $S$ & environment\footnotemark[$\dag$] & range\footnotemark[$\ddag$] & profile\footnotemark[$\S$] & $c_0$ & $a_1$ & $a_2$ & $a_3$\\
\hline 
\footnotemark[$\|$]$c_{\mathrm{V}}$ & $V_{\rm{max}} \, \rm{km \, s^{-1}}$ & global & $7\sim1500 \, \rm{km\, s^{-1}}$ & cusp & $1.75 \times 10^5$ & $-0.9$ & $0.27$ & $-0.02$\\
$c_{200}$ & $M_{200} \, M_\odot$ & global & $10^7\sim10^{15}\, M_\odot$ & cusp & $48.29$ & $-0.01863$ & $-0.008262$ & $0.0003651$\\
$c_{\mathrm{V}}$ & $V_{\rm{max}} \, \rm{km \, s^{-1}}$ & MW subhalo & $1\sim40 \, \rm{km\, s^{-1}}$ & cusp & $4.978 \times 10^4$ & $13.62$ & $-17.33$ & $5.439$\\
$c_{200}$ & $M_{200} \, M_\odot$ & MW subhalo & $10^6\sim10^{10}\, M_\odot$ & cusp & $-2.947 \times 10^2$ & $-0.4375$ & $0.05414$ & $-0.002101$\\
$c_{\mathrm{V}}$ & $V_{\rm{max}} \, \rm{km \, s^{-1}}$ & global & $7\sim1500 \, \rm{km\, s^{-1}}$ & core & $4.925 \times 10^4$ & $-0.8505$ & $0.2434$ & $-0.02314$\\
$c_{\rm{B}200}$ & $M_{200} \, M_\odot$ & global & $10^7\sim10^{15}\, M_\odot$ & core & $41.21$ & $-0.01431$ & $-0.008294$ & $0.0003589$\\
$c_{\mathrm{V}}$ & $V_{\rm{max}} \, \rm{km \, s^{-1}}$ & MW subhalo & $1\sim40 \, \rm{km\, s^{-1}}$ & core & $2.251 \times 10^4$ & $7.490$ & $-10.02$ & $3.220$\\
$c_{\rm{B}200}$ & $M_{200} \, M_\odot$ & MW subhalo & $10^6\sim10^{10}\, M_\odot$ & core & $-2.386 \times 10^2$ & $-0.4412$ & $0.05455$ & $-0.002114$\\
\hline
    \end{tabular}}\label{tab:fitting_coefficient_c-M}
\begin{tabnote}
\footnotemark[$*$]Best-ﬁt values of the parametrisations for the concentration parameter, $c_{200}$ or $c_{\rm{V}}$, as a function of $V_{\rm{max}}$ or $M_{200}$ (equation \ref{eq:cv-Vmaxrel}). 
\footnotemark[$\dag$]The third column, "environment", shows the environments the parametrisations are for. 
"MW subhalo" means that the parameters are for the subhaloes of the MW-like host.
"Global" means that the parameters are for the subhaloes without limitation on the mass of the host halo.
The "global" relation is derived from $c_{\rm{V}}$--$\vmax$ relation proposed in \citet{moline_cdm_2022}.
\footnotemark[$\ddag$]The fourth column, "range", shows the available range of these fittings.
\footnotemark[$\S$]The fifth column, "profile", shows which profile the parametrisations are for.
"Cusp" means the parametrisation is the original relation derived from the results of the cosmological $N$-body simulations and "core" means the function is derived by applying our cusp--to--core transition model to the results of the simulations (see section \ref{sec:CCtrans} for detail).\\ 
\footnotemark[$\|$]The first line shows the coefficients for the original $c_{\rm{V}}$--$\vmax$ relation in \citet{moline_cdm_2022}.
Significant figures are taken directly from \citet{moline_cdm_2022}.
Other approximation functions proposed in this paper are unified into four significant digits.
\end{tabnote}
\end{table*}

\section{Theoretical $\bar{\Sigma}(<\rmax)$--$\vmax$ relation in the Milky Way--like environment} \label{sec:MWdwarf}

Recent developments in instruments and techniques have led to remarkable progress in observing nearby galaxies. 
Moreover, near-future observations of satellite galaxies in the MW and M31 using state--of--the--art instruments (such as the Subaru Prime Focus Spectrograph on the Subaru Telescope and the James Webb Space Telescope) will provide new insights into the nature of dark matter.
This situation motivates us to explore more precisely the theoretical scaling relation for dwarf--galaxy--scale dark matter haloes in a Milky Way--like environment, analysing a data set of one of the highest--mass--resolution cosmological $N$--body simulations available described in the following subsection.
Here, we propose a new theoretical scaling relation for dwarf galaxies based on the $c$--$M$ relation of the cosmological $N$--body simulation which is of such high resolution that it allows us to study down to the scale of dwarf galaxies.

\subsection{Description of simulation data}

The cosmological $N$--body simulation, Phi--4096, which has the highest resolution among the Uchuu suite by \citet{ishiyama_Uchuu_2021}, is utilised to derive a $c$--$M$ relation for the dwarf--galaxy--mass haloes. 
Phi--4096 was conducted to extend the halo mass range down at high redshift and evaluate the resolution effects of the other simulations of the Uchuu suite.
The basic properties of Phi--4096 are listed in table \ref{tab:Phi--4096}.
Since its particle mass is $\sim 10^3\,M_\odot$, a subhalo with $10^9 \, M_\odot$ consists of $\sim 10^6$ particles.
Thanks to its relatively large volume and very high mass resolution, it has a large number of MW--like haloes and subhaloes.
Thus Phi--4096 enables the statistical study and characterization of subhaloes with a uniform resolution.
The cosmological parameters used are the matter density parameter $\Omega_0=0.31$, the baryon density parameter $\Omega_{\mathrm{b}}=0.048$, the cosmological constant $\lambda_0=0.69$, Hubble constant $h=0.68$, the spectral index of the density perturbation $n_{\mathrm{s}}=0.96$, and the matter fluctuation amplitude $\sigma_8=0.83$.
These are consistent with the observational result of the Plank Satellite \citep{planck_collaboration_planck_2020}.
Bound structures at each snapshot are identified using the \textsc{rockstar} phase space halo$/$subhalo finder \citep{behroozi_rockstar_2013}. 
The \textsc{consistent trees} code \citep{behroozi_gravitationally_2013} is used to construct merger trees.

\begin{table}
  \tbl{Properties of Phi--4096.\footnotemark[$*$] }{%
  \begin{tabular}{lcccc}
      \hline
		Name & $N$ & 
        $ \begin{array}{c} 
        L \\
        (h^{-1}\mathrm{Mpc})
        \end{array} $
        & 
        $ \begin{array}{c}
        \varepsilon \\
        (h^{-1} \mathrm{kpc})
        \end{array} $ 
        & 
        $ \begin{array}{c}
        m_\mathrm{p} \\
        (h^{-1} \mathrm{M}_{\odot})
        \end{array} $ \\
		\hline
		Phi--4096 & $4096^3$ & $16.0$ & $0.06$ & $5.13\times10^3$ \\
		\hline
    \end{tabular}}\label{tab:Phi--4096}
\begin{tabnote}
\footnotemark[$*$]
$N$, $L$, $m_\mathrm{p}$, and $\varepsilon$ are the total number of particles, box length, softening parameter, and particle mass resolution, respectively.\\ 
\end{tabnote}
\end{table}

We analyse nearly three hundred thousand subhaloes belonging to 27 MW--sized host haloes virial mass ranging from $3.40\times10^{11}$ to $2.04\times10^{12} \, h^{-1}M_\odot$ at redshift $z=0$.
The following two criteria are imposed to select the subhaloes.
First, to avoid the effects caused by the mass resolution, the virial mass needs to be larger than $10^6M_\odot$ so that a halo has at least 1,000 dark matter particles.
Second, the subhalo has to reside within twice the virial radius of a host halo at redshift $z=0$.

\subsection{$\bar{\Sigma}(<\rmax)$--$\vmax$ relation for MW satellite-like haloes}

\begin{figure}
 \begin{center}
  \includegraphics[width=\hsize]{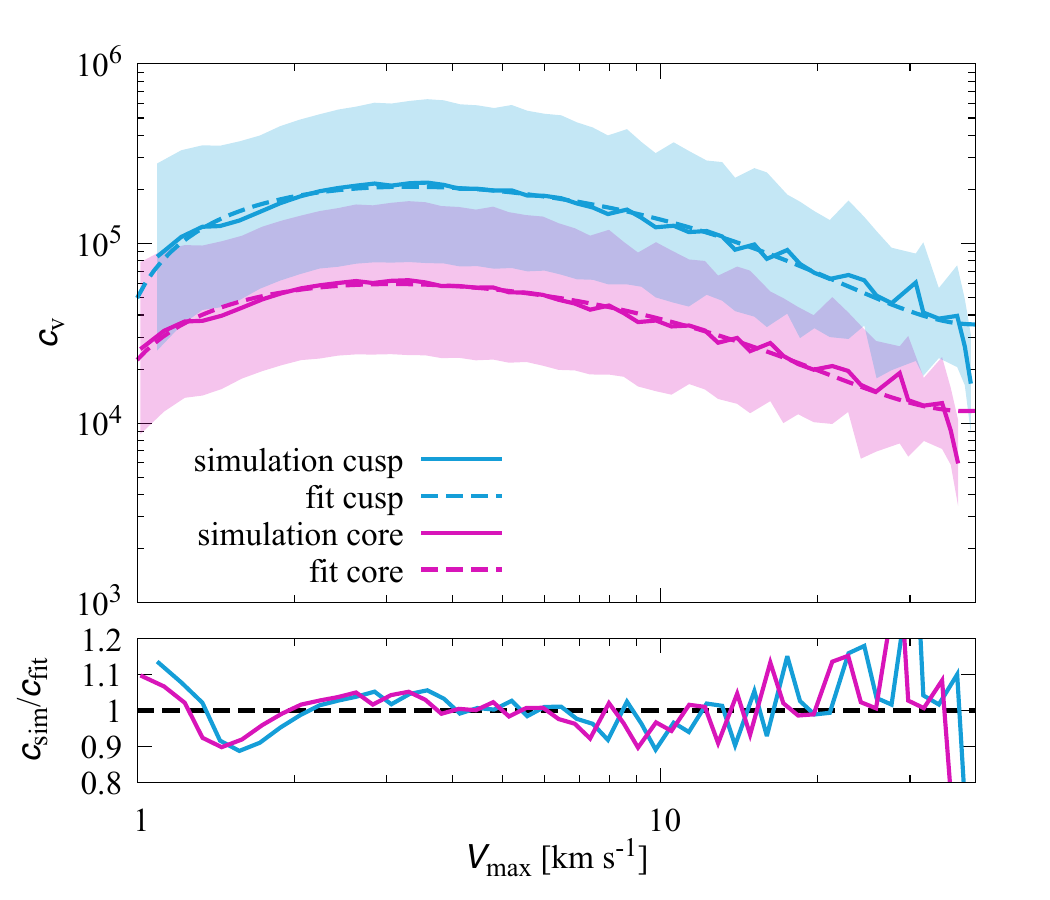}
 \end{center}
 \caption{
 Top panel: median concentration (solid lines), 25-75th percentile (shaded area), and fit (dashed lines) for a given $c_{\rm{V}}$--$\vmax$ data of subhaloes of MW--like hosts in high-resolution cosmological $N$--body simulation.
 Bottom panel: the relative difference between fit and data.
 }
\label{fig:cv-Vmaxfit}
\end{figure}

In this section, the $c$--$M$ relation derived from the subhaloes of MW--like hosts is compared with recent observations of MW satellites.
We fit the median of the concentrations inferred from the simulation by equation (\ref{eq:cv-Vmaxrel}). 
The fitting results are shown as the sky--blue solid (simulation) and dashed (fit) lines in the top panel of figure \ref{fig:cv-Vmaxfit}.
To make binned data, the exponent of the mass $M_{200}$ is divided into increments of $0.1$.
We exclude ranges containing fewer than $10$ haloes per bin.
Consequently, the data are valid for $1 < \vmax < 40$ km s$^{-1}$.
The resultant fitting function, namely the $c_{\rm{V}}$--$\vmax$ relation for subhaloes of MW--sized hosts, is expressed as in equation (\ref{eq:cv-Vmaxrel}) with $C=c_{\mathrm{V}}(V_{\max })$, $S=V_{\max } / \mathrm{km}\, \mathrm{s}^{-1}$ and
\begin{equation}
    c_0 = 4.978 \times 10^4 \; 
    \mathrm{and} 
    \; a_i=\{13.62, -17.33, 5.439\}. \label{eq:cv-Vmaxrel_MWsat}
\end{equation}
The sky--blue line in the bottom panel of figure \ref{fig:cv-Vmaxfit} illustrates the relative error in the fitting function for the cusp profile. 
It demonstrates that the error remains below 10 per cent with $\vmax$ between $1.5\,\mathrm{km\, s}^{-1}$ and $15\,\mathrm{km\,s}^{-1}$, indicating that is a reliable approximation within this range.
Furthermore, we fit the median $c_{200}$ as a function of $M_{200}$ through equation (\ref{eq:cv-Vmaxrel}) with $C=c_{200}(M_{200})$, $S=M_{200} / M_\odot$ and find the coefficients to be
\begin{equation}
    c_0 = -2.947 \times 10^2 \; 
    \mathrm{and} 
    \; a_i=\{-0.4375, 0.05414, -0.002101\}. \label{eq:c200-M200rel_MW}
\end{equation}
These coefficients are organised in table \ref{tab:fitting_coefficient_c-M}.

We adopted the observational estimation of the density profile of satellites of MW and M31 derived by \citet{hayashi_universal_2017}.
\citet{hayashi_universal_2017} estimated the dark matter halo profile with a higher precision than previous studies by fitting the axisymmetric Jeans equation to the line--of--sight velocities obtained from spectroscopic observations.
They consider the following density profile
\begin{eqnarray}
    \rho(R, z) &=& \rho_0\left(\frac{m}{b_{\mathrm{halo}}}\right)^\alpha\left[1+\left(\frac{m}{b_{\mathrm{halo}}}\right)^2\right]^{-(\alpha+3) / 2}, \label{eq:rho_axi}\\
    m^2 &=& R^2+z^2 / Q^2, \label{eq:rho_axi_m}
\end{eqnarray}
where $\rho_0$ is a scale density, $b_\mathrm{halo}$ is a scale length in the spatial distribution, and $Q$ is an axial ratio for the dark matter halo.

The comparison between the $\bar{\Sigma}(<\rmax)$--$\vmax$ relation of MW-satellite-like haloes and observation of MW and M31 satellites is shown in figure \ref{fig:scalings}b.
The solid grey line describes the median of $\bar{\Sigma}(<\rmax)$ of MW-satellite-like haloes in the cosmological $N$-body simulation Phi-4096 and the shaded region illustrates the range between the 25th percentile and the 75th percentile.
The black dotted curve is the theoretical $\bar{\Sigma}(<\rmax)$ curve drawn on the basis of the $c_{\rm{V}}$--$\vmax$ relation (equation \ref{eq:cv-Vmaxrel} and \ref{eq:cv-Vmaxrel_MWsat}) substituted in equation (\ref{eq:Sigmavmax-Vmax_relation}).
Note that the applicable range for the theoretical $\bar{\Sigma}(<\rmax)$--$\vmax$ relation, derived from our $c$--$M$ relation, is $1 < \vmax < 40$ km s$^{-1}$, which corresponds to the typical range for subhaloes.

Figure \ref{fig:scalings}b shows that the red symbols indicating the eight MW classical dwarfs and five M31 dwarfs measured by \citet{hayashi_universal_2017} are in excellent agreement with this theoretical $\bar{\Sigma}(<\rmax)$--$\vmax$ relation.
Because of the limited number and significant errors of the observed dwarf galaxies in the previous studies, the characteristic mean surface density inside $\rmax$ has been proposed to be constant. 
However, the theoretical $\bar{\Sigma}(<\rmax)$--$\vmax$ relation we derived here is consistent with recent observations of MW dwarf galaxies, showing that the average surface density inside $\rmax$ depends on the system size.
We expect future observation using state-of-the-art telescopes and instruments to estimate the density distributions of haloes of satellite galaxies with $\vmax < 20 \, \mathrm{km\, s^{-1}}$.

\section{Theoretical scaling relation from dwarf galaxies to galaxy clusters} \label{sec:wide}

In Section \ref{sec:scaling}, we compared observations of the $\bar{\Sigma}(<\rmax)$--$\vmax$ relations on galaxy scale with that derived from the $c$--$M$ relation predicted by the CDM model. 
The result clearly shows that the observed scaling relations are deeply rooted in the CDM prediction.
In this section, we extend the theoretical scaling relation not only to $\vmax-\bar{\Sigma}(<\rmax)$ plane but also to other physical quantities. 
Then, these relations are compared with the observed quantities for a wide range of astronomical objects, from dwarf galaxies to galaxy clusters.

We adopt the fitting function (equations \ref{eq:cv-Vmaxrel} and \ref{eq:cv-Vmaxrel_Moline2023}) proposed by \citet{moline_cdm_2022}.
While the fitting function of theoretical $c$--$M$ relation we propose (equation \ref{eq:cv-Vmaxrel_MWsat}) is applicable for subhaloes associated with the MW--sized hosts ($10^{6}\sim10^{10}M_\odot$), the fitting function suggested by \citet{moline_cdm_2022} is available for broader mass range $10^{7}\sim10^{14}M_\odot$.
To compare the theoretical $c$--$M$ relation with observations ranging from dwarf galaxies to galaxy clusters, we adopt the fitting function proposed by \citet{moline_cdm_2022} in the following subsections.

\subsection{Description of employed observational data} \label{sec:obsdata}
This section describes employed observational data and selection criteria adopted to ensure enough data quality.

We adopted the parameters which are derived assuming that haloes have an NFW profile in order to simply compare them with the parameters of haloes inferred from cosmological $N$--body simulations.

For dwarf galaxies, we employ data from \citet{oh_high-resolution_2015}.
\citet{oh_high-resolution_2015} derived rotation curves from “Local Irregulars That Trace Luminosity Extremes, The \ce{H_I} Nearby Galaxy Survey” (LITTLE THINGS) and constructed mass models of 26 galaxies.
17 $c_{200}$ and $V_{200}$ on their table 2 are used.

For massive galaxies, we adopt the data from rotation curves obtained with spectroscopic observations by \citet{spano_ghasp_2008}, \citet{de_blok_high-resolution_2008}, and \citet{sofue_rotation_2016}.
\citet{spano_ghasp_2008} derived the mass models of 36 spiral galaxies using rotation curve analysis.
They combine the \ce{H_{\alpha}} 2D velocity field obtained using Fabry–Perot interferometry as part of the GHASP survey and \ce{H_I} data from the literature.
$\rN$ (columun 10) and $\rhoN$ (columun 11) of their table 2 are used.
\citet{de_blok_high-resolution_2008} presented the mass models of 19 galaxies using rotation curve analysis from The \ce{H_I} Nearby Galaxy Survey (THINGS).
$c_{200}$ (columun 8) and $V_{200}$ (columun 9) of their table 6 are used.
\citet{sofue_rotation_2016} compiled rotation curves of more than one hundred spiral galaxies from the literature.
$\rN$ (columun 7), $r_{200}$ (columun 9), and $M_{200}$ (columun 10) of their table 1 are used.

Groups and clusters of galaxies are shown using the data from \citet{gastaldello_probing_2007}, \citet{merten_clash_2015},  and \citet{umetsu_clash_2016}.
\citet{gastaldello_probing_2007} presented radial mass profiles for 16 relaxed galaxy groups.
Using the X-ray data from Chandra and XMM-Newton, they solved the equation of hydrostatic equilibrium to calculate the gravitating mass distribution.
$\rN$ (columun 3) of their table 3, $c_{200}$ (columun 8), and $M_{200}$ (columun 10) of their table 7 are used.
\citet{merten_clash_2015} have determined the mass distribution of $X$--ray selected galaxy clusters from the Cluster Lensing and Supernova Survey with Hubble (CLASH), spanning a redshift range between $0.19$ and $0.89$.  
Data from HST are used to analyse both weak and strong gravitational lensing and data from ground--based telescopes, primarily Subaru Telescope, are utilised for weak--lensing analysis.
$\rhoN$ (columun 2) and $\rN$ (columun 3) of their table 6,  and $M_{200}$ (columun 6) and $c_{200}$ (columun 7) of their table 7 are used.
\citet{umetsu_clash_2016} provided comprehensive analysis of strong--lensing, weak--lensing shear and magnification data for a sample of $X$--ray--regular and high--magnification galaxy clusters at $0.19 \lesssim z \lesssim 0.69$ selected from CLASH.
They constructed mass models of individual clusters from a joint analysis of the full lensing constraints.
$M_{200}$ (columun 2), $c_{200}$ (columun 3), and $\rN$ (columun 4) of their table 2 are used.
The object analysed in \citet{umetsu_clash_2016} overlaps with the objects analysed in \citet{merten_clash_2015}. 
We adopt the updated values from \citet{umetsu_clash_2016} whenever possible.

To ensure data quality, objects with unreliable estimation are excluded from our sample.
The exclusion criteria are as follows:
\begin{enumerate}
    \item Data points exhibiting uncertainties larger than the expected values are excluded.
    \footnote{
This criterion excludes the data of \citet{salucci_dwarf_2012} and \citet{hayashi_universal_2017} already shown in figure \ref{fig:scalings}.
}
    \item \citet{spano_ghasp_2008} introduced artificial thresholds for the stellar mass-to-luminosity ratios of the disc and bulge, $\rN$, and $\rhoN$, to eliminate non-physical values. We exclude data affected by these thresholds.
\end{enumerate}

\subsection{Comparison between theoretical scaling relations and observations}

\begin{figure*}
    \begin{center}
    \includegraphics[width=\hsize]{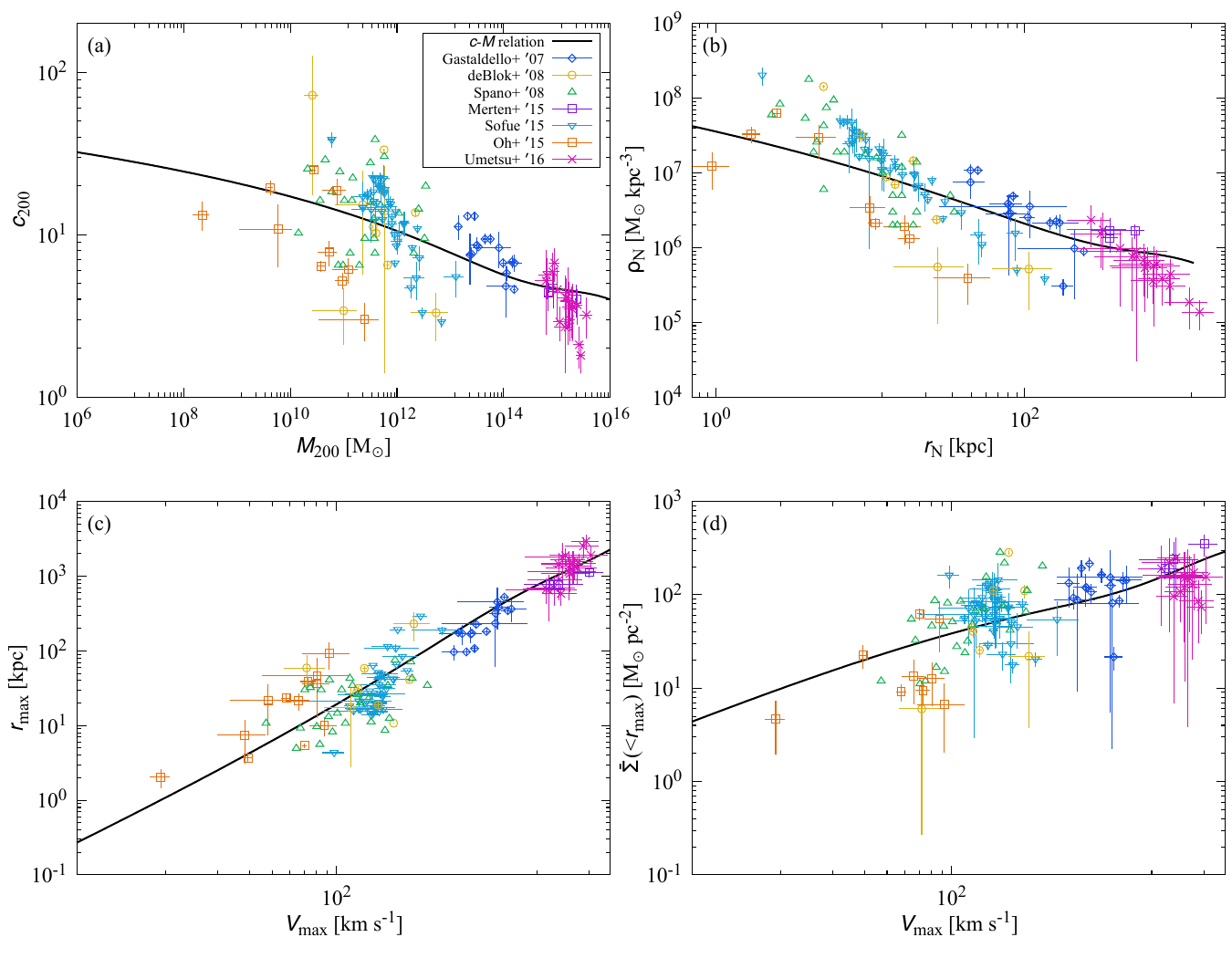}
    \end{center}
    \caption{Comparisons between observations (symbols) and (a) the $c$--$M$ relation, (b) $\rN$--$\rhoN$ relation, (c) $\vmax$--$\rmax$ relation, (d) $\bar{\Sigma}(<\rmax)$--$\vmax$ relation. 
    The orange symbols represent nearby dwarf galaxies from \citet{oh_high-resolution_2015}.
    The yellow, green, sky--blue symbols show bright galaxies by \citet{de_blok_high-resolution_2008}, \citet{spano_ghasp_2008}, and \citet{sofue_rotation_2016}.
    The blue symbols are groups of galaxies from \citet{gastaldello_probing_2007}.
    The purple and magenta symbols correspond to clusters of galaxies from \citet{merten_clash_2015} and \citet{umetsu_clash_2016}, respectively.
    }
    \label{fig:obs}
\end{figure*}
  
Here, the theoretical scaling relations based on the $c$--$M$ relation of the CDM model are compared with the observational data for four combinations of parameters.

In figure \ref{fig:obs}a, the $c$--$M$ relation is compared with the observation from dwarfs to clusters.
The vertical axis shows $c_{200}$, and the horizontal axis shows $M_{200}$.
The black solid curve shows the fitting function proposed by \citet{moline_cdm_2022} (equation \ref{eq:cv-Vmaxrel} and \ref{eq:c200-M200_rel_wide}), and the symbols represent the observed values of $c_{200}$ and $M_{200}$.
The theoretical $c_{200}$ slowly decreases as the mass increases.
The theoretical $c$--$M$ relation is consistent with the observations in the whole mass range.
However, a significant deviation is seen for the observed dwarf galaxies.
It should be noted that there is a significant ambiguity in the mass determination of dwarf galaxies because the velocity dispersion is usually comparable level to the rotation velocity in such systems.
Therefore, spectroscopic observation with higher resolution and dispersion and the development of mass models are expected to constrain the $c$--$M$ relation models more accurately.

Observation and theory are compared on the $\rN$--$\rhoN$ plane in figure \ref{fig:obs}b.
To convert $c_{200}$ and $M_{200}$ to $\rN$ and $\rhoN$, we use the following equations: 
\begin{equation}
    \rhoN = \frac{200\rhocrit c_{200}^3}{3 \fN(c_{200})},
\end{equation}
and
\begin{equation}
    \rN = r_{200} / c_{200},
\end{equation}
where
\begin{equation}
    r_{200} = \left(\frac{3}{4}\frac{M_{200}}{200\pi\rhocrit}\right)^{1/3}.
\end{equation}
The observations indicate that $\rhoN$ decreases as $\rN$ increases in agreement with the theoretical line.
Dwarf galaxies, which have the $\rN$ of several $\mathrm{kpc}$, spread around the theoretical curve due to the errors of their mass distribution.
The spiral galaxies are in good accordance with the theoretical prediction, with radii from $1$ to $100 \, \mathrm{kpc}$ and the scale densities range from $10^6$ to $10^8 \, M_\odot \, \mathrm{kpc^{-3}}$.
On the other hand, a small discrepancy is seen in the low mass and massive ends.
One possible reason is that some assumptions needed to estimate the mass of the stellar component, such as maximum/minimum disk models or a mass--to--light ratio, to decompose the rotation curve have a systematic effect on the parameters of the haloes.
The scale radii of galaxy clusters range from hundreds of $\mathrm{kpc}$ to several thousands of $\mathrm{kpc}$ and the scale densities range from $10^5$ to $10^7M_\odot\,\mathrm{kpc^{-3}}$.

Figure \ref{fig:obs}c shows the comparison on the $\vmax$--$\rmax$ plane.
The vertical axis indicates $\rmax$, and the horizontal axis corresponds to $\vmax$.

We convert $\rN$ and $\rhoN$ values to $\vmax$ and $\rmax$ with equation (\ref{eq:rmaxN}) and 
\begin{equation}
    \vmax = \sqrt{4\pi G \rhoN {\rN}^2 \frac{\fN(\xN)}{x_\mathrm{N}}}.
\end{equation}
The theoretical curve shows that $\rmax$ increases as $\vmax$ increases. 
For the lower mass range, $\rmax$ is approximately proportional to $\vmax$ on a log-log scale.
The theoretical relation nicely reproduces the observed values for two orders of magnitude in the maximum circular velocity or seven orders of magnitude in the virial mass.

Figure \ref{fig:obs}d shows $\bar{\Sigma}(<\rmax)$ as a function of $\vmax$ as in figure \ref{fig:scalings}.
The theoretical relation nicely reproduces the observations.
According to the theoretical $\bar{\Sigma}(<\rmax)$--$\vmax$ relation, more massive objects have larger characteristic mean surface density.

On all planes, there is a certain amount of dispersion on the observational data set for the range of the galaxies and dwarf galaxies.
This may be caused by using NFW profiles when deriving the dark matter halo profiles, rather than using a profile with a core more suitable to reproduce the observed profiles (see section \ref{sec:intro}).
Until now, we adopted the NFW profile fit for comparing with the results of cosmological $N$--body simulations.
In the next section, we extend the $c$--$M$ relation to cored profiles and compare with observations fitted with cored profiles.

\section{Theoretical scaling relation after the cusp--to--core transition} \label{sec:CCtrans}

In this section, we construct a scaling relation for cored haloes based on the $c$--$M$ relation.
In previous sections, we compared $\Lambda$CDM model predictions with observations assuming the dark matter density profile to have a central cusp.
However, observations of disc galaxies, dwarf galaxies, and low surface brightness disc galaxies generally indicate the existence of a core in the centre of the dark matter haloes (see section \ref{sec:intro}).
Most scaling relations hold for the cored density distribution. 
While several solutions to solve the cusp--core problem are introduced in section \ref{sec:intro}, we especially focus on the baryonic solution in this section.
We stand on the hypothesis that dark matter haloes formed primarily with a cuspy density distribution and then some dynamical processes lead to the formation of a core at the centre of the dark matter haloes.
We simplify the complicated physical processes and provide a model to convert the parameters of a cuspy profile, namely, the scale density and the scale radius, to the parameters of a core profile, the central density and the core radius.

\subsection{Cusp--to--core transition model} \label{subsec:CCtransmodel}

In this section, our cusp--to--core transition model is described.
We adopt the Burkert profile as a core profile and the NFW profile as a cusp profile.
In order to enhance applicability, we deliberately omit specific physical processes responsible for the cusp--to--core transition when constructing a model.
\citet{ogiya_connection_2014} proposed a crude transition model that can be analytically solved, assuming mass conservation and that only the inner part is changed during the cusp--to--core transition, considering $r_{200} \gg \rN, \, r_\mathrm{B}$.
They had $\rB = \eta \rN$ and $\rhoB = \zeta \rhoN$ where $\eta \approx 1.24$ and $\zeta \approx 0.526$ and regard these as $\eta \approx 1$ and $\zeta \approx 1$.
However, in practice, $c_{200} = r_{200}/r_\mathrm{N}$ is on the order of ten and cannot be regarded as infinity.
Therefore, the actual coefficients $\eta$ and $\zeta$ depend on $c_{200}$ and a more careful approach is required.

We impose the same assumptions as \citet{ogiya_connection_2014}, however, adopt more sophisticated conditions to formulate a transition model.
First, only the density distribution around the centre is affected, and the one in the outskirts does not change.
Second, the virial mass $M_{200}$ of a dark matter halo is conserved.
From the first condition, equating the density distribution of the NFW profile and the Burkert profile at $r=r_{200}$, we have
\begin{equation}
    \frac{\rho_\mathrm{N}}{\frac{r_{200}}{r_\mathrm{N}}\left(\frac{r_{200}}{r_\mathrm{N}}+1\right)^2} = \frac{\rho_\mathrm{B}}{\left(\frac{r_{200}}{r_\mathrm{B}}+1\right)\left(\frac{r_{200}^2}{{r_\mathrm{B}}^2}+1\right)}  \label{eq:(i)}
\end{equation}
As a requirement that meets the second condition, we adopt the conservation of virial masses, in other words, we equate equations(\ref{eq:m_nfw}) and (\ref{eq:m_burkert}) at $r=r_{200}$ and yield
\begin{equation}
    \rho_\mathrm{N} r_\mathrm{N}^3 f_\mathrm{N}\left(\frac{r_{200}}{r_\mathrm{N}}\right) = \rho_\mathrm{B} r_\mathrm{B}^{3} f_\mathrm{B}\left(\frac{r_{200}}{r_\mathrm{B}}\right). \label{eq:(ii)}
\end{equation}
Eliminating $\rho_\mathrm{N}$ or $\rhoB$ from equations (\ref{eq:(i)}) and (\ref{eq:(ii)}) and substituting $r_\mathrm{B} = \eta r_\mathrm{N}$, we obtain
\begin{equation}
    g(\eta) - \frac{ f_\mathrm{N}(c_{200})}{ f_\mathrm{B}(c_{200}/\eta)} = 0,  \label{eq:bisection_method}
\end{equation}
where 
\begin{equation}
    g(\eta) \equiv \frac{(c_{200}+\eta)(c_{200}^2+\eta^2)}{c_{200}(c_{200}+1)^2}.
\end{equation}

We solve equation (\ref{eq:bisection_method}) for $\eta$ numerically at each $c_{200}$ value of dark matter haloes.
Here, note that $r_{200}$ is determined from a given combination of $r_\mathrm{N}$ and $\rho_\mathrm{N}$ using equation (\ref{eq:m_nfw}) and (\ref{eq:m200}).
Then substituting $r_\mathrm{N}$ into equation (\ref{eq:(i)}) or equation (\ref{eq:(ii)}), we obtain $\rho_\mathrm{N}$.

In this way, we can derive the value of $\rho_\mathrm{N}$ and $r_\mathrm{N}$ from a given combination of $\rho_\mathrm{B}$ and $r_\mathrm{B}$.

\subsection{Theoretical scaling relations for cored haloes} \label{sec:core_relation}

Combining the prediction for the cuspy haloes inferred from the cosmological $N$--body simulation with our cusp--to--core transition model, we can predict the theoretical relation for dark matter haloes which have experienced cusp--to--core transitions. 
Here, we convert the properties of cuspy haloes described by the theoretical $c$--$M$ relation to the properties of cored haloes using our cusp--to--core transition model.
Specifically, we derive $\rhoN$ and $\rN$ from the theoretical $c$--$M$ relation and convert them into $\rhoB$ and $\rB$ using our cusp--to--core transition model.
Then we recover $c_{\rm{V}}$ and $\vmax$ from $\rhoB$ and $\rB$ and fit them with equation (\ref{eq:cv-Vmaxrel}).
The fitting function of the properties of cored haloes after this transition, in other words, the theoretical $c_{\rm{V}}$--$\vmax$ relation for cored haloes from dwarf galaxies to clusters of galaxies is given by equation (\ref{eq:cv-Vmaxrel}) with $C=c_{\mathrm{V}}(V_{\max })$, $S=V_{\max } / \mathrm{km}\, \mathrm{s}^{-1}$ and
\begin{equation}
    c_0 = 4.925 \times 10^4 \quad 
    \mathrm{and}
    \quad a_i=\{-0.8505, 0.2434, -0.02314\}. \label{eq:cv-Vmaxrel_core_wide}
\end{equation}
We also perform a fitting for the median of $c_{200}$ against $M_{200}$ for cored haloes from dwarf galaxies to clusters of galaxies.
The concentration for the Burkert profile is defined by
\begin{equation}
    c_{\rm{B}200} = \frac{r_{200}}{r_{\rm{B}}}.
\end{equation}
Adopting equation (\ref{eq:cv-Vmaxrel}) as fitting function with $C=c_{200}(M_{200})$ and $S=M_{200} / M_\odot$, coefficients become
\begin{equation}
    c_0 = 41.21 \; 
    \mathrm{and} 
    \; a_i=\{-0.01431, -0.008294, 0.0003589\}.
\end{equation}
The error of this parameterization is less than 3 per cent. 
The $c_{\rm{V}}$--$\vmax$ relation for cored haloes of dwarf galaxies of MW--sized hosts is given by equation (\ref{eq:cv-Vmaxrel}) with $C=c_{\mathrm{V}}(V_{\max })$, $S=V_{\max } / \mathrm{km}\, \mathrm{s}^{-1}$ and
\begin{equation}
    c_0 = 2.251 \times 10^4 \; 
    \mathrm{and} 
    \; a_i=\{7.490, -10.02, 3.220\}. \label{eq:cv-Vmaxrel_core_MWsat}
\end{equation}
The fitting result is shown in the magenta line in figure \ref{fig:cv-Vmaxfit}.
We also perform a fitting for the median of $c_{200}$ against $M_{200}$ for cored haloes of dwarf galaxies.
Assuming equation (\ref{eq:cv-Vmaxrel}) for fitting function with $C=c_{200}(M_{200})$ and $S=M_{200} / M_\odot$, coefficients become
\begin{equation}
    c_0 = -2.386 \times 10^2 \; 
    \mathrm{and} 
    \; a_i=\{-0.4412, 0.05455, -0.002114\}. \label{eq:c200-M200rel_MW_core}
\end{equation}
These coefficients are also listed in table \ref{tab:fitting_coefficient_c-M} for a quick look.

\begin{figure*}
    \begin{center}
        \includegraphics[width=\hsize]{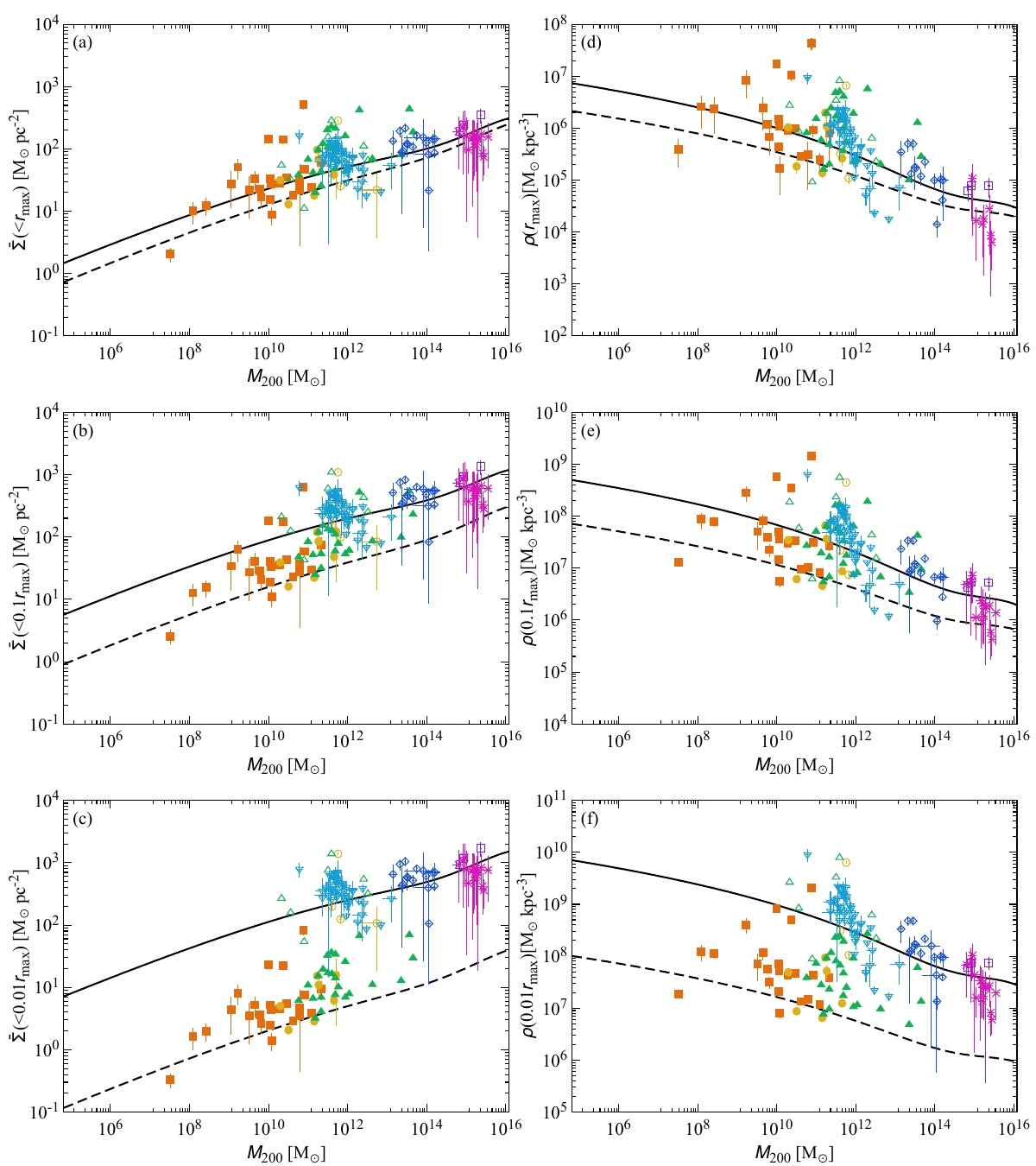}
    \end{center}
    \caption{The left--hand column shows characteristic mean surface density inside $X$ times $\rmax$ against $M_{200}$.
    The right--hand column shows three--dimensional density at $X$ times $\rmax$ against $M_{200}$.
    The theoretical scaling relation for the NFW dark matter profile is shown as the solid lines (equivalent to those used in figure \ref{fig:obs}).
    The theoretical scaling relation after the cusp--to--core transition is shown as the dashed line.
    The symbols represent the estimated properties of dark matter haloes associated with dwarf galaxies, galaxies, groups and clusters of galaxies by observation.
    Objects for which a core profile is adopted are plotted as filled symbols, and for which a cusp profile is adopted are plotted as open symbols. 
    }
    \label{fig:CCtrans}
\end{figure*}

We compare the $c$--$M$ relation for cuspy haloes and cored haloes in figure \ref{fig:CCtrans}.
In the left column, the characteristic mean surface density inside a fraction of $\rmax$ against $M_{200}$ is shown.
The characteristic mean surface density can be defined for both a cusp and a core profile, while the central surface density can only be defined for core profiles.
Moreover, the characteristic mean surface density can be easily inferred from observables compared to the inner slope of the dark matter profile since it requires only the determination of the mass inside a certain radius and does not need the recovery of a full-density profile.
These enable us of robust discussions using the characteristic mean surface density.
The right column shows three--dimensional density at a fraction of $\rmax$ against $M_{200}$.
The theoretical scaling relation for cuspy haloes is shown as the solid line.
The theoretical scaling relation for core haloes after the cusp--to--core transition is shown as the dashed lines.
The top two panels of figure \ref{fig:CCtrans} show $\bar{\Sigma}(<\rmax)$ against $M_{200}$ and $\rho(\rmax)$ against $M_{200}$, respectively.
The difference between the theoretical scaling relation of the cuspy profile and the one for the cored profile is pretty small.
To tell the difference between the cusp and the core profile, we must see the inner density distribution since the density distributions in the outer part are almost the same.
The two lower rows in figure \ref{fig:CCtrans} are similar to the upper row but at radii $X \rmax$ with $X<1$.
The smaller the $X$ value is, the larger the difference between the density of a cuspy halo and a cored halo becomes.
Then cusps and the cores can be distinguished using $\bar{\Sigma}(<X \rmax)$ or $\rho(X \rmax)$.
Although we examined various $X$, the results reported here are only for $X = 1, 0.1$ and $0.01$.

The top panels in figure \ref{fig:CCtrans} indicate that if the spatial resolution of mass determination by observations is equal to or worse than $\rmax$, it is impossible to distinguish between the core profile and the cusp profile of the dark matter halo.
In the middle panels, the density at $X=0.1$ has a difference of one order of magnitude between the theoretical prediction for the cusps and the cores, hence, should be easy to distinguish observationally.
The density at $X=0.01$ has a difference of two orders of magnitude between the theoretical prediction for the cusps and the cores.

The characteristic mean surface density $\bar{\Sigma}(<Xr_{\rm{max}})$ and the density $\rho(Xr_{\rm{max}})$ can be derived from the $c$--$M$ relation (equations \ref{eq:cv-Vmaxrel} and \ref{eq:c200-M200_rel_wide}). 
However, for convenience, it is desirable to provide an approximation that describes the dependence of the characteristic mean surface density and the density on the $X$ and the mass of the dark matter halo. 
We propose the fitting function for the characteristic mean density and the density:
\begin{equation}
    \log_{10}[D(M_{200}, X)] = \log_{10}[F(X)] + \sum_{i=0}^4 a_i \left[\log_{10}{\left(\frac{M_{200}}{M_\odot}\right)}\right]^i.\label{eq:sigma_rho_fitting}
\end{equation}
Here, $D(M_{200}, X)$ indicates the characteristic mean density or the density at $X r_{\rm{max}}$ and $F(X)$ undertakes the $X$ dependency. $D(M_{200}, X)$, $F(X)$ and coefficients $a_i$ are listed in table \ref{tab:fitting_coefficient_Sigma_rho}. 
These ﬁts work well in the halo mass $10^7 M_\odot < M_{200} < 10^{15} M_\odot$, its accuracy is better than 11 per cent at all $M_{200}$ values within this range and $X = \{1, 0.1, 0.01\}$.

\begin{table*}
  \tbl{Coefficients of equation (\ref{eq:sigma_rho_fitting}).\footnotemark[$*$]}{%
  \begin{tabular}{cccccccc}
      \hline
	$D$ & profile\footnotemark[$\dag$] & $F$ &$a_0$ & $a_1$ & $a_2$ & $a_3$ & $a_4$ \\
      \hline
	$\bar{\Sigma}(<Xr_{\rm{max}}) \, M_\odot \, \rm{pc}^{-2}$ & cusp & $\frac{1}{X^2} \frac{\fN(X \xN)}{\fN(\xN)}$ & $0.1491$ & $-0.3566$ & $0.1122$ & $-0.008745$ & $0.0002310$\\
	$\bar{\Sigma}(<Xr_{\rm{max}}) \, M_\odot \, \rm{pc}^{-2}$ & core & ${\frac{1}{X^2} \frac{\fB(X \xB)}{\fB(\xB)}}$ & $-0.5081$ & $-0.1924$ & $0.08468$ & $-0.006754$ & $0.0001825$\\
	$\rho(<Xr_{\rm{max}}) \, M_\odot \, \rm{kpc}^{-3}$ & cusp & $\frac{(\xN + 1)^2}{X (X\xN + 1)^2}$ & $9.635$ & $-1.211$ & $0.1975$ & $-0.01525$ & $0.0003990$\\
	$\rho(<Xr_{\rm{max}}) \, M_\odot \, \rm{kpc}^{-3}$ & core & $\frac{(\xB + 1)(\xB^2 + 1)}{(X\xB + 1)(X^2\xB^2 + 1)}$ & $8.4911$ & $-0.9302$ & $0.1505$ & $-0.01187$ & $0.0003174$\\
      \hline
  \end{tabular}
  }\label{tab:fitting_coefficient_Sigma_rho}
\begin{tabnote}
\footnotemark[$*$]
Best-ﬁtting parameters of the characteristic mean surface density-mass and the density-mass relation for the model equation (\ref{eq:sigma_rho_fitting}) are summarised.
\footnotemark[$\dag$]The second column, "profile", shows which profile the parametrisations are for.
"Cusp" means the parametrisation is the original relation derived from the results of the cosmological $N$-body simulations and "core" means the function is derived by applying our cusp--to--core transition model to the results of the simulations (see section \ref{sec:CCtrans} for detail).
\end{tabnote}
\end{table*}

\subsection{Observational verification of the cusp--to--core transition} \label{sec:CCprob}

The observational data points are added to the theoretical prediction in figure \ref{fig:CCtrans}.
The symbols represent the estimated properties of dark matter haloes associated with dwarf galaxies, galaxies, groups and clusters of galaxies by observations.
The colour assignments are the same as in the previous sections.
For papers that published both cusp profile and core profile fitting results and $\chi^2$-value for each fitting \citep{de_blok_high-resolution_2008, spano_ghasp_2008}, the fitting result with the better $\chi^2$-value is adopted.
\citet{oh_high-resolution_2015} listed both cusp profile and core profile fitting results, however, did not list $\chi^2$-value.
They mentioned that the decomposed dark matter rotation curves of most sample galaxies are well matched in shape to those of core--like haloes.
Therefore, we adopt core profile fits for all galaxies.
For objects that are fitted by the p--ISO, we convert the p--ISO parameters to the Burkert profile parameters using the method described in Appendix \ref{app:p-ISO_to_Burkert}.
In this way, it becomes possible to compare the observations with the parameters after adopting our cusp--to--core transition model, which assumes the Burkert profile as a cored profile.
Objects for which the core profiles are adopted are plotted as filled symbols, and for which the cusp profile is adopted are plotted as open symbols. 

Here, we can obtain hints of the system size dependence of the cusp--to--core transition from figure \ref{fig:CCtrans}.
In the bottom two panels, less massive objects, such as dwarf galaxies and galaxies are located between the $c$--$M$ relation of the cusp profile and the core profile,  while massive objects, such as clusters of galaxies are well placed over the $c$--$M$ relation of the cusp.
This indicates that massive galaxies and galaxy clusters did not undergo cusp--to--core transitions and remain cuspy.
Moreover, it is seen that the transition mass between the dark matter halo with a cusp and halo with a core is at about $10^{11}\,M_\odot$, with the transition range extending from  $10^{10}\,M_\odot$ to $10^{12}\,M_\odot$ on either side.
This is consistent with the statement of the classical model advocated by \citet{dekel_origin_1986} that $10^{11}\,M_\odot$ is the mass at which the energy that cumulative supernovae inject into the surrounding gas and the energy required to carry the gas to infinity against the gravitational potential of the dark matter halo are comparable.
In the less massive region ($<10^{10}\,M_\odot$), there exist both dark matter haloes with a cusp and that with a core.
This indicates that among less massive galaxies, some objects experience significant cusp--to--core transitions while others weakly experience it or do not.
Similar trends have also been seen in recent galaxy formation simulations. 
For example, NIHAO \citep{tollet_nihao_2016} and FIRE-2 \citep{lazar_dark_2020} have shown that for objects that have smaller stellar--to--halo mass ratio, the mass distribution near the centre of the dark matter halo turns from cusp to core.
What is even more interesting is that these simulations demonstrate that dark matter haloes in faint dwarf galaxies are little affected by stellar feedback and that the mass distributions in the central parts of dark matter haloes retain cuspy structures (\cite{tollet_nihao_2016}; \cite{lazar_dark_2020}). 
On the other hand, analysis of faint dwarf galaxies around the MW by \authorcite{hayashi_diversity_2020} (\yearcite{hayashi_diversity_2020}, \yearcite{hayashi_dark_2022}) shows a mixture of cusp--like dark matter haloes and core--like dark matter haloes in those galaxies. 
We will postpone discussing the properties of dark matter haloes of ultra--faint dwarf galaxies to our upcoming paper.

For MW subhaloes, the 2.5th to 97.5th (the 25th to 75th) percentile range of theoretical $\bar{\Sigma}(<X\rmax)-M_{200}$ relation for cuspy haloes and cored haloes do not overlap when $X<0.021$ ($X<0.15$). 
If the characteristic mean surface density can be measured for $X < 0.021$ ($X < 0.15$), it will be possible to determine the critical transition mass at which the cusp--to--core transition occurs in dark matter haloes.
For example, if there is a $10^8 \, M_\odot$ object ($\rmax \sim 0.6 \, \mathrm{kpc}$ from the $c$--$M$ relation for MW--satellite--like haloes; equation \ref{eq:cv-Vmaxrel_core_MWsat}), $X \rmax $ corresponds to $0.01 \rm{kpc}$ for $X=0.021$ ($0.09 \rm{kpc}$ for $X=0.15$).
To resolve $0.01 \rm{kpc}$ ($0.09 \rm{kpc}$) at a distance of $10 \rm{Mpc}$ (local universe), the required angular resolution is at least $0.21$ arcseconds ($1.9$ arcseconds).
Better than angular resolution $\sim 0.21$ arcseconds is expected to be achievable with the Atacama Large Millimeter/submillimeter Array (ALMA), in the near future, the Next Generation Very Large Array (ngVLA), the Thirty Meter Telescope (TMT), and other telescopes.

In this paper, the least massive sample of a galaxy we have is $\sim 10^8 \, M_\odot$.
Nearby ultra--faint dwarf galaxies are suitable to test our theoretical prediction on the less massive side ($< 10^8 \, M_\odot$).
Ultra--faint dwarf galaxies offer a more accurate determination of the central density distribution in dark matter haloes because they are uninfluenced by baryonic effects.
By measuring the spatial distribution of stellar velocity dispersion and combining it with dynamical equilibrium models, one can derive the dark matter density profiles.
The latest and upcoming telescopes and observational devices, for example, Subaru Prime Focus Spectrograph (PFS) is anticipated to achieve an angular resolution $\sim30$ arcseconds at its peak performance.
If $X=0.021$ ($X=0.15$) is adopted as the criterion for determining the critical transition mass, dwarf galaxies with dark matter halo masses exceeding $10^8M_\odot$ within 90 kpc (600 kpc) from us are considered favourable targets for this telescope.

\section{Discussions} \label{sec:discussion}
In this section, we discuss the functional form of the dark matter halo profiles used in the observations, and the possible extension of our cusp--to--core transition model.

\subsection{Functional form of the dark matter halo profile}

The archival observation data used in our analysis assumed the cuspy functional form or the cored functional forms of mass density distributions. 
Actually, there are other profiles suggested to describe the observational mass distribution of haloes, however, studying them goes beyond the scope of this paper. 
Further study to include a function with varying inner slope, such as the Einasto profile, is needed.

\subsection{Application of the cusp--to--core transition model}

\subsubsection{Inner slope of the density and transition mass}
Our cusp--to--core transition model assumes either a cusp or a flat core.  
In other words, expressing the mass density distribution near the centre as $\rho(r)\propto r^\gamma$, only $\gamma = -1$ or $\gamma = 0$ are considered.   
In reality, however, there is a possibility that a cusp may change to a shallower cusp or moderate core ($-1<\gamma<0$).
The $c$--$M$ relation of moderately transformed halo should be between the $c$--$M$ relation for the cusp profile (solid line) and the one for the core profile (dashed line) in figure \ref{fig:CCtrans}.
Actually, the observed galaxies lie between the prediction for the cuspy haloes and for the cores around $10^{11}\, M_\odot$ that is the mass thought to be at which the energy that cumulative supernovae inject into the surrounding gas and the energy required to carry the gas to infinity against the gravitational potential of the dark matter halo are comparable \citep{dekel_origin_1986}.
These galaxies might have intermediate profiles 
($-1<\gamma<0$) which we did not take into account in this paper.
It is not clear how the deviation caused by forcibly fitting a halo with slopes of $-1$ or $0$ affects the conclusions in this paper.

\subsubsection{Stability of a core}
The stability of a core is closely connected with the balance between the time scale for the relaxation of a dark matter halo and the duration of intensive star formation.
We assume that the halo retains a core profile once a halo experiences the cusp--to--core transition.
This assumption is verified if below two conditions are satisfied.

Firstly, the cusp--to--core transition should occur after the halo reaches the equilibrium state.
\citet{onorbe_forged_2015} showed that a core formed in the early epoch can be erased as a halo continues to accrete matter and experience central mergers using zoom--in cosmological hydrodynamic simulations. 

Secondary, stellar feedback should occur recurrently.
\citet{ogiya_core-cusp_2011} numerically experimented the dynamical response of dark matter haloes to the variance of the gravitational potential induced by instantaneous gas removal from galaxy centres.
They found the removal of gas can decrease the inner density, however, after a few dozen dynamical times, a shallow cusp recovers.
On the other hand, \citet{ogiya_core-cusp_2014} found that if there is a recurrent change of gravitational potential driven by the cycle of star formation and stellar feedback, the core retains for a few dozen dynamical times.
Many previous hydrodynamic simulation studies reveal periodic star formation in the early epoch of galaxy formation, supporting a recurrent change of gravitational potential.
Consequently, it is reasonable to assume the core retains for a few dozen dynamical times.

\subsubsection{Epoch of the cusp--to--core transition}

While we transform the scaling relation for cuspy haloes at $z=0$ to construct a scaling relation for cored haloes, the actual cusp--to--core transition should have taken place in the past.
The epoch of the cusp--to--core transition depends on the formation epoch of the dark matter halo and the star formation history.
Since the star formation history is strongly affected by stellar feedbacks and radiative feedbacks, the ambiguity of the star formation model becomes very large depending on the accuracy of the stellar feedback model and the radiative feedback model. 
Therefore, as the first step, our study simply applies the cusp--to--core transition model to the $c$--$M$ relation at $z=0$.  
As a next step, it is interesting to consider the epoch of the cusp-to-core transition as a function of redshift, taking into account the formation of the dark matter halo and the star formation history \citep{ogiya_connection_2014}.
The validity of our model can be discussed if we compare these results with observations of high-redshift galaxies.

\section{Conclusion} \label{sec:conclusion}

In this paper, we investigate the theoretical and observational scaling relations from dwarf galaxies to clusters of galaxies.
We summarise our main findings as:
\begin{enumerate}
\item 
The observational scaling relations between properties of galaxy-sized dark matter haloes \citep{burkert_structure_1995, spano_ghasp_2008, donato_constant_2009, salucci_dwarf_2012, kormendy_scaling_2016} are consistent with the theoretical scaling relation based on the $c$--$M$ relation of dark matter haloes formed in cosmological $N$--body simulations.
In other words, the observed scaling relations seem to originate from the $c$--$M$ relation of the CDM haloes.

\item 
Utilising the results of Phi--4096, a cosmological $N$--body simulation whose resolution is high enough to enable us to discuss the properties of MW satellite--sized haloes, we derived the $c$--$M$ relation for the MW--satellite sized subhaloes at $z=0$.
The fitting functions are shown in equation (\ref{eq:cv-Vmaxrel}) with equation (\ref{eq:cv-Vmaxrel_MWsat}) or (\ref{eq:c200-M200rel_MW}).

\item 
The $c$--$M$ relation derived from the Uchuu suite by \citet{moline_cdm_2022} reproduces observations going from dwarf galaxies to clusters of galaxies in four parameters planes, the $c_{200}$--$M_{200}$ plane, the $\rN$--$\rhoN$ plane, the $\vmax$--$\rmax$ plane, and the $\vmax$--$\bar{\Sigma}(\rmax)$ plane.
However, the dispersion of parameters is broad from dwarf galaxies to massive galaxies.
This might be because of forcibly fitting the NFW profile to the cored haloes.

\item 
We provide the scaling relation for cored haloes for a wide mass range from dwarf galaxies to clusters as equation (\ref{eq:cv-Vmaxrel}) with equation (\ref{eq:cv-Vmaxrel_core_wide}) and for the MW's satellite mass range as equation (\ref{eq:cv-Vmaxrel}) with equation (\ref{eq:cv-Vmaxrel_core_MWsat}), converting the $c$--$M$ relation for a cuspy profile to the one for a cored profile using the cusp--to--core transition model proposed in this work.

\item 
Comparing the theoretical scaling relation for cored haloes with observations, we find:
a) Less massive objects, such as dwarf galaxies and galaxies lie over both the theoretical scaling relation for cored haloes and cuspy haloes, while massive objects, i.e., groups of galaxies and clusters of galaxies are well placed over the theoretical scaling relation for cuspy haloes.
There is a critical mass around $10^{11}\,M_\odot$ for the transition between the dark matter halo with a cusp and that with a core.
This may be consistent with the statement of \citet{dekel_origin_1986} that $10^{11}\,M_\odot$ is the mass at which the stellar feedback and the gravitational potential of the dark matter halo are comparable.
This indicates that on smaller scales, some objects experience strong cusp--to--core transitions while others 
weakly experience it or do not.
On the other hand, massive objects do not undergo cusp--to--core transitions and remain cuspy.
b) If the characteristic mean surface density for $X < 0.021$ (to discriminate 95 per cent of dark matter halos formed in cosmological $N$-body simulation) or $X < 0.15$ (50 per cent) is determined, it is possible to identify the critical transition mass at which the cusp--to--core transition of the dark matter halo occurs.
The latest and upcoming telescopes and observational devices, such as Subaru PFS, ALMA, TMT, and ngVLA would achieve this resolution for the nearby dwarf galaxies.
\end{enumerate}


\begin{ack}
We thank the anonymous referee for the detailed review and comments, and Tomoaki Ishiyama for providing the results of the cosmological $N$--body simulation Phi--4096.
We are grateful for the helpful comments from Kohei Hayashi and Masashi Chiba regarding observational aspects,  especially about local dwarf galaxies, and for valuable perspectives on future observations from Takuya Hashimoto. We are grateful to Yudai Kazuno for many essential discussions and Moemi Tanuma for providing valuable material. This work was supported in part by JSPS KAKENHI Grant Number JP24K00669, JP24K07085, JP23KJ0280, JP22KJ0370, and JP20K04022,  and by Multidisciplinary Cooperative Research Program in CCS, University of Tsukuba.
\end{ack}

\appendix
\section*{Matching the p-ISO to the Burkert profile} \label{app:p-ISO_to_Burkert}
In this section, how the parameters of the p-ISO are altered to the parameters of the Burkert profile is explained.
In section \ref{sec:core_relation}, we provide a prediction for cored halo with the Burkert profile in the figure \ref{fig:CCtrans}.
On the other hand, some observational studies modelled dark matter haloes as the p-ISO. 
However, our cusp--to--core transition model cannot be applied to the p-ISO since it is proportional to $r^{-2}$ at the outer region while the NFW profile is proportional to $r^{-3}$.
In other words, the core radius and the central densities of the p-ISO which satisfy two requirements of our cusp--to--core transition model do not exist.
Thereby, we scale the parameters of the p-ISO to those of the Burkert profile.
We describe how we translate $r_\mathrm{p-ISO}$ and $\rho_\mathrm{p-ISO}$ into $r_\mathrm{B}$ and $\rho_\mathrm{B}$ below.

Since the accuracy of observations decreases as we see a more distant radius from the centre of a galaxy, reproducing the density distribution of the central region is essential.
Therefore, we consider the central density expressed as a parameter of the p-ISO to be identical to the central density of the Burkert profile:
\begin{equation}
    \rho_\mathrm{p-ISO} = \rho_\mathrm{B}.
\end{equation}
Then, we expect that masses enclosed within the core radius are the same. Equating equation (\ref{eq:m_burkert}) and (\ref{eq:m_p-ISO}) at core radius of the p-ISO, and supposing $r_\mathrm{B} = \xi r_\mathrm{p-ISO}$, we have
\begin{equation}
    r_\mathrm{p-ISO}^{3} f_\mathrm{p-ISO}(1) - r_\mathrm{B}^{3} f_\mathrm{B}\left(\frac{1}{\xi}\right) = 0. \label{eq:Burkert_to_p-ISO}
\end{equation}
Solving equation (\ref{eq:Burkert_to_p-ISO}) for $\xi$ by numerical procedure, we obtain $r_\mathrm{B}$ from given $r_\mathrm{p-ISO}$.

\bibliographystyle{pasj}
\bibliography{main_v3.1}

\end{document}